# High-speed impulsive stimulated Brillouin microscopy

**Jiarui Li,[1] Taoran Le,[1] Hongyuan Zhang,[2] Haoyun Wei,[1] and Yan Li[1, *]**

[1]*State Key Laboratory of Precision Measurement Technology and Instruments, Department of Precision Instrument, Tsinghua University, Beijing 100084, China*
[2]*Cole Eye Institute, Cleveland Clinic, Cleveland 44195, USA*
*Corresponding author: liyan@mail.tsinghua.edu.cn*

**Abstract:** Brillouin microscopy, which maps elastic modulus from the frequency shift of scattered light, has evolved to a faster speed for the investigation of rapid biomechanical changes. Impulsive stimulated Brillouin scattering (ISBS) spectroscopy has the potential to speed up measurement through the resonant amplification interaction from pulsed excitation and time-domain continuous detection. However, significant progress has not been achieved due to the limitation in signal-to-noise ratio (SNR) and the corresponding need for excessive averaging to maintain high spectral precision. Moreover, the limited spatial resolution also hinders its application in mechanical imaging. Here, by scrutinizing the SNR model, we design a high-speed ISBS microscope through multi-parameter optimization including phase, reference power, and acquisition time. Leveraging this, with the further assistance of the Matrix Pencil method for data processing, three-dimensional mechanical images are mapped under multiple contrast mechanisms for a millimeter-scale polydimethylsiloxane pattern immersed in methanol, enabling the identification of these two transparent materials without any contact or labeling. Our experimental results demonstrate the capability to maintain high spectral precision and resolution at a sub-millisecond integration time for one pixel. With a two-order improvement in the speed and a tenfold improvement in the spatial resolution over the state-of-the-art systems, this method makes it possible for ISBS microscopes to sensitively investigate rapid mechanical changes in time and space.

## 1. Introduction

Mechanical properties play a critical role in connecting the mechanics of the microenvironment to the behavior of cells and tissues, allowing for the translation of mechanical stimuli into biochemical signals [1-3]. While there is significant progress in our understanding of biochemical signaling, the study of biomechanical interactions has been hindered by limited methods with a high spatial resolution and without contact. Based on the interaction of light and acoustic phonons, Brillouin microscopy maps the spatial distribution of elastic modulus by deriving the frequency information of scattered light, providing a non-contact and label-free approach to reveal what viscoelasticity is encountered in different regions at high resolution [4, 5]. This emerging technique has made initial contributions to ocular biomechanics [6-8], tumor biology [9], and developmental biology [10, 11]. The demand for faster Brillouin microscopy in biomechanics has become increasingly urgent, particularly in volumetric imaging and the analysis of rapid processes, making fast measurement a focal issue [9, 11, 12]. However, the measurement speed of existing instruments still needs improvement.

Despite significant advancements in the applications of biomechanical imaging [5, 13-15] and in vivo diagnostics [8, 16, 17] over the past decade, confocal spontaneous Brillouin microscopy has almost reached the speed limit (20 ms/spectrum) due to the weak spontaneous interaction process. Many efforts have been made to overcome this speed limitation. In terms of optical architecture, the line-scanning Brillouin microscopy enables the interaction between light and spontaneous acoustic phonons in a line and the simultaneous detection of several hundred pixels in the vertical direction [9, 11, 12], achieving a measurement speed of 1 ms per Brillouin spectrum, but with limited spectral resolution. In terms of excitation principles, stimulated Brillouin microscopy with a pump-probe scheme efficiently excites resonant phonons [10, 18, 19], yet its potential for speed enhancement is limited by the insufficient utilization of nonlinear interaction by continuous sources (CW) (5 ms/spectrum). In this case, a pulsed nonlinear scattering scheme will address these issues.

Impulsive stimulated Brillouin scattering (ISBS) microscopy [20-22], assisted with pulsed excitation beams, has the potential to obtain a Brillouin spectrum in a single shot. A pair of pump pulses excite the acoustic wave, the CW probe is scattered in the focal volume, and the scattered light describes the acoustic oscillations in the time domain. Assisted with another CW reference beam, the heterodyne detection architecture allows for enhanced efficiency. After transforming this time-domain heterodyne signal through the fast Fourier transformation (FFT) and spectral fitting, the elasticity and viscosity of the material in the focal volume can be derived from the frequency shift and linewidth of the Brillouin peak, respectively. However, due to the low signal-to-noise ratio (SNR) of a single-shot excitation, the measurement speed of ISBS is hindered by the need for multiple averages, even with sufficient source power. To speed up the measurement, the Matrix Pencil (MP) method, a time-domain spectral analysis method, is introduced to ISBS spectrometers for a higher noise suppression ability and fewer averaging times [23], but the application of this system is still limited to liquid samples due to its high photodamage, and the restricted spatial resolution prevents it from achieving microscopic imaging. Higher spatial resolution often corresponds to a larger divergence angle, and the maximum scattering efficiency can only be achieved when the incident angle of the probe is at the Bragg angle, where the probe, scattered light, and acoustic field satisfy the momentum matching condition. Therefore, increasing spatial resolution comes with a decrease in ISBS efficiency. To achieve higher spatial resolution for three-dimensional microscopic ISBS imaging while maintaining spectral precision, further optimization of the SNR is necessary.

In this paper, we develop an SNR enhancement solution for time-domain Brillouin signals, with the assistance of an SNR model and corresponding multi-parameter optimization. An ISBS microscope is accordingly designed to deliver a two-order enhancement in speed, a tenfold increase in spatial resolution, and reduced photodamage compared to state-of-the-art ISBS microscopes. We perform optimized phase compensation, reference power ratio, and acquisition time, achieving a 15.6 dB SNR for the Brillouin spectrum of a hydrogel sample under an integration time of 0.3 ms and total optical energy of tens of microjoules, while maintaining a spectral precision of 1 MHz with the assistance of the MP method. Furthermore, this time-domain data processing method is immune to spectral broadening and sidebands caused by acquisition time optimization, thus preserving the high spectral resolution advantage of ISBS. On this basis, three-dimensional imaging, with a spatial resolution of 13×14×160 μm$^3$, is performed to characterize the various viscoelastic properties of a transparent millimeter-scale polydimethylsiloxane (PDMS) sample without contact or labeling, advancing a high-speed ISBS microimaging.

## 2. Principle, theory, and verification

*2.1 ISBS and mechanical properties*

In the stimulated Brillouin scattering process with pump and probe beams, the phonons experience resonant amplification. The coupling of the optical and acoustic fields can generate scattering signals stronger than the spontaneous process and essentially unaffected by the elastic background. Based on this nonlinear process, ISBS employs a pair of ultra-short pulses to pump acoustic waves and a CW laser to probe. Due to the broadband feature of pump pulses, various frequency differences exist between the two intersecting pump beams. According to the conservation of energy and momentum, this allows resonance with multiple phonon modes and excites a pair of counter-propagating acoustic waves. The combination of a transmission grating (TG) with a period of $d_T$ and a 4f system with an amplification rate of $f_2/f_1$, as shown in Fig. 1, stabilizes the wavelength of the acoustic wave $\Lambda$ at $f_2 d_T/(2f_1)$. It also results in an acoustic frequency, $v_a = V_a/\Lambda$, that remains unaffected by the laser frequency, normal refractive index, and density of the material, where $V_a$ is the sound speed. Through Brillouin scattering, a CW beam probes the refractive index changes induced by acoustic waves. Benefiting from the TG-4f geometry, the probe beam automatically incidents at the Bragg angle which maximizes the scattering efficiency. The probe beam, which has a narrow spectral linewidth, interacts with the excited acoustic waves through Stokes and anti-Stokes scattering, resulting in a frequency shift. After detecting the oscillation of this scattered light, whose amplitude is modulated, the Brillouin spectrum can be obtained by applying Fourier transformation to the time-domain signal.

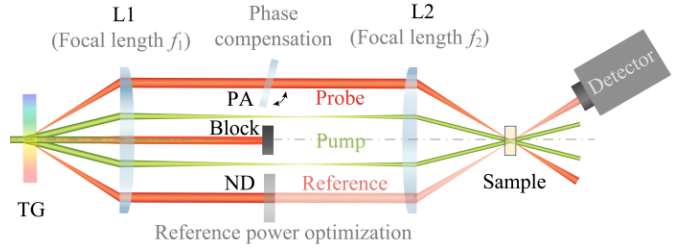

Fig. 1. Schematic of ISBS with heterodyne detection. TG: transmission grating, L1 and L2: lenses, PA: phase adjustment plate.

According to the detection scheme for scattered light, current ISBS spectrometers can be classified into non-heterodyne and heterodyne configurations. In non-heterodyne mode, the scattered light of the probe is directly detected, resulting in a measured Brillouin frequency shift $v_B$ corresponding to the beat frequency of the Stokes and anti-Stokes light, which is $2v_a$. While in heterodyne mode, the reference beam, attenuated by another CW beam symmetrical to the probe beam, transmits through the sample without the acoustic modulation and overlaps with the scattered light of the probe beam, resulting in interference. Due to the reference beam at the same frequency as the probe, the measured frequency $v_B$ becomes the beat frequency between the reference beam and the Stokes or anti-Stokes light, coinciding with the acoustic wave frequency $v_a$. This enhances the Brillouin signal and provides a direct reflection of the material's elastic properties. The corresponding spectrum to this heterodyne signal is [24]

$$Sp(v) \propto \frac{d_z \rho \frac{\partial n}{\partial \rho} \sqrt{I_{probe} I_{ref}} \cos\varphi}{\lambda_{probe} \cos\theta_B} \frac{A}{\sqrt{(\alpha V_a)^2 + [2\pi(v - v_a)]^2}}, \quad (1)$$

where $d_z$ is the axial thickness of the acoustic perturbation region, $\rho \partial n/\partial \rho$ is the elasto-optic coefficient of the material, $I_{probe}$, $\lambda_{probe}$, and $\theta_B$ are the intensity, wavelength, and Bragg angle of the probe, respectively, $I_{ref}$ represents the intensity of the reference beam, $\varphi$ is the phase difference between the probe and reference, and $A$ and $\alpha$ represent the amplitude and attenuation coefficient of the acoustic wave, respectively. Therefore, the speed and attenuation of the acoustic wave can be obtained based on the measured frequency shift $v_B$ and linewidth $\Delta_B$ from the heterodyne ISBS spectrometer, which is given by

$$V = \frac{f_2 d_T}{2 f_1} v_B$$
$$\alpha = \frac{2\pi f_1}{\sqrt{3} f_2 d_T} \frac{\Delta_B}{v_B} \quad . \quad (2)$$

The spectral characteristics of Brillouin-scattered light reflect the propagation properties of acoustic waves in the material, which in turn are related to the material's viscoelasticity. Based on the spectral information, the real and imaginary parts of the complex longitudinal modulus of the material can be given by [25]

$$M' = \rho V^2 = \rho(\frac{f_2 d_T}{2f_1})^2 v_B^2$$
$$M'' = \rho V^3 \frac{\alpha}{\pi v_B} = \frac{\rho}{\sqrt{3}}(\frac{f_2 d_T}{2f_1})^2 v_B \Delta_B \quad , \quad (3)$$

where $\rho$ represents the mass density. In ISBS, the complex longitudinal modulus of materials can be derived only based on density and spectral information without the refractive index, but density measurement also complicates the process. The introduction of Brillouin loss tangent (BLT) can simplify the characterization of local viscoelasticity, defined as the ratio of the imaginary part to the real part of the complex longitudinal modulus [26], which is

$$\tan(\varphi_M) = \frac{M''}{M'} = \frac{\Delta_B}{\sqrt{3} v_B} \quad . \quad (4)$$

Combining BLT with ISBS microscopy makes it possible to assess the spatial distribution and temporal changes of viscoelasticity in a simpler way, solely based on Brillouin spectral information.

*2.2 SNR model of heterodyne ISBS*

In ISBS spectroscopy, increasing the SNR of the signal obtained from a single pulse excitation allows high spectral precision with fewer averages, thereby speeding up the measurement. By establishing a theoretical model for SNR, we can analyze the key parameters and accordingly design the ISBS system to achieve the optimal SNR. According to Eq. (1), the reference beam assists in amplifying weak signals and reducing common-mode noise, but its effect on improving SNR also depends on the noise characteristics of the system. The main sources of noise for a heterodyne ISBS spectrometer are dark current noise, shot noise, and thermoelectrical noise, all of which exhibit characteristics of white noise in the spectral domain. Among them, dark current noise and thermoelectrical noise are both independent of the incident light power and can be simplified as a constant background noise $i_b$ for the same detection system. On the other hand, shot noise $i_s$ is dependent on the incident light power and can be described as $\sqrt{2q\Delta f I_{PD}}$, where $q$ is the electron charge, $\Delta f$ is the bandwidth of the detection, and $I_{PD}$ is the average photocurrent and is proportional to the incident optical power when operating in the linear region. In this case, the standard deviation of noise in the heterodyne ISBS system is proportional to $\sqrt{i_b^2 + r_s R_{\text{ref}}}$, where $R_{\text{ref}}$ represents the power ratio of the reference to the probe, and $r_s$ is the simplified proportionality factor of shot noise. Combining Eq. (1) with the above analysis, the spectral SNR of the heterodyne ISBS spectrometer is determined by

$$SNR \propto \frac{A d_z \rho \frac{\partial n}{\partial \rho} I_{\text{probe}}}{\alpha V_a \lambda_{\text{probe}} \cos \theta_B} \frac{\sqrt{R_{\text{ref}}} \cos \varphi}{\sqrt{i_b^2 + r_s R_{\text{ref}}}} \quad . \quad (5)$$

On this basis, the spectral SNR can be optimized in terms of the following aspects.

*2.2.1 Phase mismatch and compensation*

In the heterodyne detection scheme, either the attenuation plate in the reference beam or the tilting of the sample may introduce an additional phase difference $\varphi$ between the reference and the probe beam, which affects the spectral peak in the form of a cosine. To ensure temporal coherence and reduce the impact of phase mismatch on the SNR, a phase adjustment flat (PA) is introduced in the probe beam, compensating for the phase difference between the two optical paths, as illustrated in Fig. 1. After each installation of the sample or adjustment of the attenuation plate in the reference beam, PA is rotated to introduce an additional phase to the probe beam and compensate for the temporal incoherence. The impact of PA angle on the spectral peak can be simply represented as $\cos[\varphi_0+2\pi(n_{PA}-1)d_{PA}/(\lambda_{probe}\cos\theta)]$, where $\varphi_0$ represents the initial phase difference between the probe and the reference, while $n_{PA}$ and $d_{PA}$ denote the refractive index and axial thickness of PA, respectively, and $\theta$ is the angle between PA and the normal to the optical axis.

To validate this theory, ISBS signals of ethanol are measured at 41 different PA angles with an interval of 1°. In Fig. 2(a), the spectral peak varies with angle, showing that the spectral peak reaches its local maximum value at multiple angles, consistent with the theoretical expectations. According to Eq. (5), since the noise is mainly influenced by the power of the reference beam and is hardly affected by the weak heterodyne signals of the probe and reference, the addition of phase compensation avoids the SNR loss due to phase mismatch.

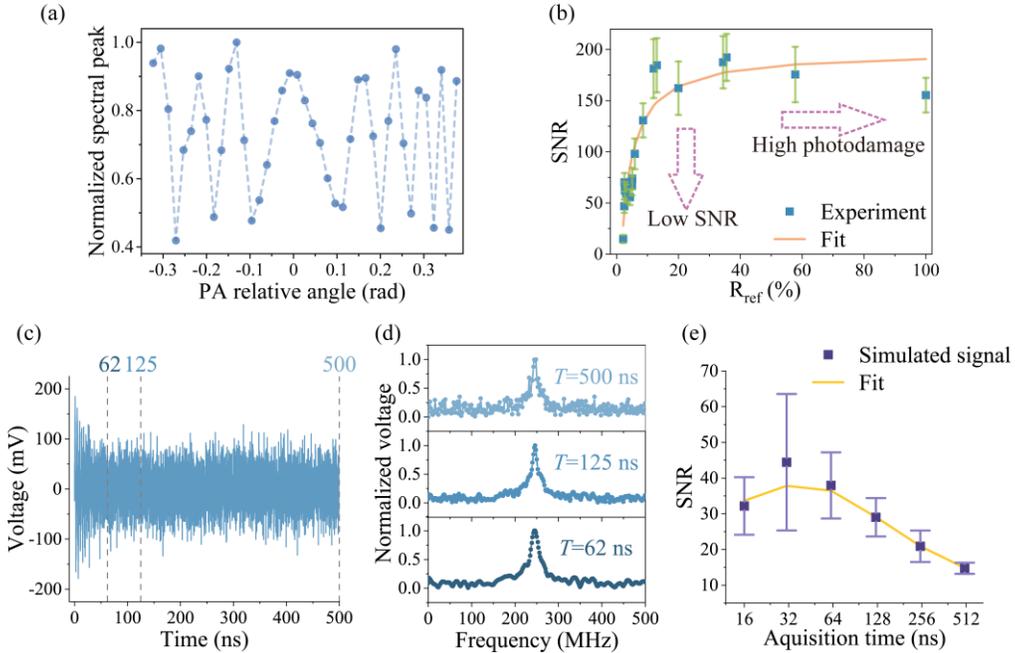

Fig. 2. The key parameters affecting the SNR. (a) The experimental spectral peak for ethanol spectra at different PA angles. (b) Experimental SNR and theoretical fitting for methanol spectra acquired with varying power ratios $R_{ref}$ between the reference and the probe. (c) Simulated time-domain signal of PDMS and (d) corresponding spectra at three acquisition times. (e) Spectral SNRs of simulated signals at various acquisition times and the theoretical fitting results.

### 2.2.2 Reference power consideration

As indicated by Eq. (5), increasing the power ratio $R_{ref}$ between the reference beam and the probe beam can contribute to the improvement of the SNR. However, under a constant CW source power, when $R_{ref}$ increases to the point where shot noise $i_s$ becomes significantly larger than background noise $i_b$, there exists an SNR limit because both signal and noise increase with the square root of $R_{ref}$. In this case, further increasing $R_{ref}$ will not enhance the SNR. Excessive average power not only limits the SNR due to detector saturation but also elevates the risk of thermal damage to the sample. Therefore, in heterodyne ISBS systems, there exists an optimal power ratio $R_{ref}$ of the reference beam.

To verify this theory, methanol signals are measured at various $R_{ref}$, and the corresponding spectral SNRs are calculated. Each measurement is performed with an integration time of 26 ms. The two pump beams deliver a total of 13 µJ of single-pulse energy to the sample at a repetition rate of 10 kHz, while the probe and unattenuated reference beams each have a power of 9 mW. By replacing neutral density filters (ND) with different transmittances, the spectral SNRs are obtained at different $R_{ref}$, with 50 repeated measurements at each $R_{ref}$. After each ND replacement, the PA angle is adjusted to eliminate the impact of phase mismatch on the SNR. In Fig. 2(b), the blue squares represent the averaged spectral SNRs at each $R_{ref}$, and the corresponding error bars indicate the standard deviations. Based on Eq. (5), a least-squares fitting is performed on the experimental data, with an R-square of 0.9, and the fitting curve is represented by the yellow line in Fig. 2(b), further confirming the reliability of the theoretical SNR model. For this measurement condition, the estimated SNR limit is around 23 dB by applying fitting parameters to the SNR model. To determine the optimal $R_{ref}$ for different measurement conditions, the trade-off between SNR improvement and the risk of photodamage must be considered.

### 2.2.3 Impact of acquisition time

Since ISBS is a time-domain acquisition method, adjusting the acquisition parameters allows for a convenient optimization of the spectral quality. In the analysis above, an ideal sampling situation with infinite duration is assumed, while in the experimental situation, the acquisition time duration $T$ is finite, which also affects the ISBS spectral SNR. On one hand, for white noise, which is the primary type of noise source, its power spectral density remains constant. So in the case of finite time-domain acquisition, the collected noise energy increases linearly with the acquisition time $T$. On the other hand, ISBS spectra are amplitude spectra, so the spectral noise is affected by the acquisition time in a square root relationship. Fig. 2(c) simulates a time-domain signal of PDMS, and Fig. 2(d) shows the corresponding spectra at three different acquisition times. When the acquisition time is reduced from 500 ns to 62 ns, the spectral SNR increases from 17 to 34, resulting in a two-fold improvement.

However, this does not imply that a smaller acquisition time always leads to a better SNR. There exists an optimal range of acquisition time that maximizes the SNR. When $T$ is too small to obtain the complete signal, the spectral peak decreases as $T$ decreases, and this relationship is proportional to $[1-\exp(-\alpha VT)]/(\alpha V)$ [27]. Therefore, the spectral SNR varies with the acquisition time $T$ as

$$SNR(T) \propto \frac{1-\exp(-\alpha VT)}{\alpha V \sqrt{T}} \tag{6}$$

To validate this theory, the time-domain signal of PDMS is simulated 10 times, and the spectral SNRs at different acquisition times are calculated for each simulation. The average and standard deviation of the SNR at each acquisition time are shown as purple rectangles and error bars in Fig. 2(e). Based on Eq. (6), the mean SNR values corresponding to different $T$ are fitted, the fitting result of which is shown as a yellow curve in Fig. 2(e) with an R-square of 0.99, confirming the reliability of this theory. In experimental measurements, optimization of the acquisition time should be considered based on the different damping characteristics of the target material.

## 3. Methods

### 3.1 Experimental setup

For three-dimensional mechanical imaging, an ISBS microscope is constructed with improved spatial resolution, lower photodamage, and enhanced SNR [Fig. 3(a)], which is integrated into a microimaging structure. The pulsed pump laser (Huaray Laser, PINE-532-15) at 532 nm, with tunable repetition rate (set as 100 kHz) and 10 ps pulse duration, is focused on the TG (HOLO

OR, DS-278-Q-Y-A, with a grating period of 20.7 μm) using a cylindrical lens CL (Thorlabs, LJ1144RM-A, with a focal length of 150 mm). After passing through L3 (Thorlabs, AC254-150-A, with a focal length of 150 mm) and L4 (Thorlabs, AC254-060-A, with a focal length of 60 mm), the pump laser is then focused on the sample with a volume of 472 μm × 16 μm × 7.3 mm [Fig. 3(b)] which is measured by a beam quality analyzer (Spiricon, BGS-USB-SP620). The focused beam maintains a lateral range that contains approximately 120 acoustic wave spatial periods (acoustic waves with a wavelength of 4.1 μm) while increasing the longitudinal energy density to ensure efficient excitation of acoustic waves. The CW laser (Thorlabs, DBR780PN, and New Focus, TA-7613) at 780 nm is focused on TG using a spherical lens L2 with a focal length of 125 mm and then divided by TG into the probe beam and the reference beam. This configuration, combined with L3 and L4, achieves a spatial resolution of 13×14×160 μm$^3$ on the sample [Fig. 3(b)], enabling three-dimensional imaging.

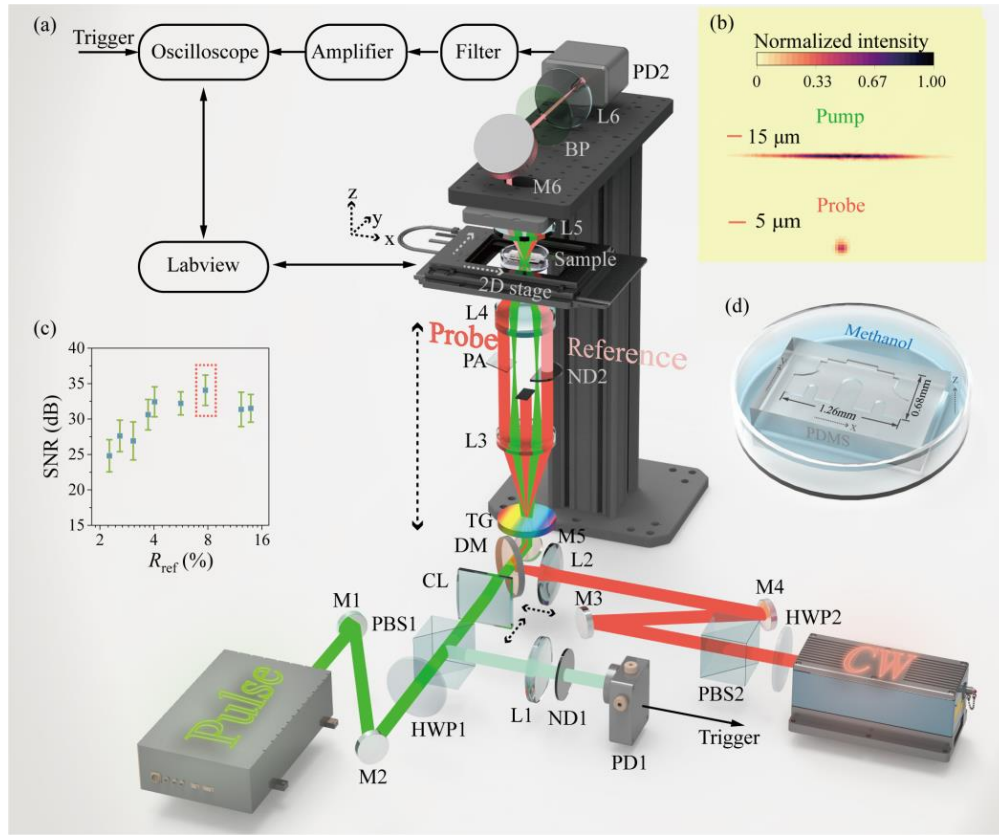

Fig. 3. (a) The setup of the heterodyne ISBS microscope. M1-M6: mirror, HWP1 and HWP2: half-wave plate, PBS1 and PBS2: polarizing beam splitter, L1-L6: spherical lens, CL: cylindrical lens, ND1 and ND2: neutral density filters, DM: dichroic mirror, TG: transmission grating, PA: phase adjustment plate, BP: 780 nm bandpass filters, PD1 and PD2: photodetector. (b) The spot size of the pump and the probe on the focal plane. (c) $R_{ref}$ optimization at the set measurement parameters for imaging. (d) Sample schematic.

Before the imaging scan, the PA angle is adjusted to achieve the maximum amplitude of the time-domain signal through phase compensation. The heterodyne ISBS signal is detected by a photodetector PD2 (New Focus, 1601FS-AC, 30 kHz-1 GHz), and then passes through a high-pass filter (Mini Circuits, 133 MHz-1 GHz) and two-stage cascade amplifiers (Mini Circuits, ZFL-1000+, 1 GHz). Taking into account the influence of $R_{ref}$ on SNR and the saturation of PD2 [Fig. 3(c)], an attenuation group ND2 with an optical density (OD) of 1.3 is used to achieve the optimal SNR for PDMS. The time-domain signal is acquired and stored with an acquisition time of 66.9 ns using a high-sampling-rate oscilloscope (Agilent Technologies, DSO9254A, 2.5 GHz), and the 30-time averaging process of the signal is also performed on the oscilloscope. The enhanced SNR allows for a spectral integration time as low as 0.3 ms, with a total single-pulse energy of 1.5 µJ for the pump and a CW laser power of 60 mW and 5 mW for the probe and the reference on the sample, respectively. During the integration time, the total energy density within the focal volume on the sample is 0.001 nJ/µm$^3$ at 532 nm and 1 nJ/µm$^3$ at 780 nm, without causing obvious photodamage on PDMS.

For three-dimensional imaging, the microscope is equipped with a fast scanning stage (Thorlabs, MLS203-1) to hold and scan the sample in the x-y plane. The scanning in the z-direction is achieved by a motorized translation stage (Thorlabs, ZFM2020) to move the TG, L3, and L4 together, and the corresponding focusing of CL and L2 is accomplished using two additional translation stages. The high-speed translation stage and oscilloscope are synchronized and controlled using home-built LabVIEW scripts, enabling automated acquisition and storage of ISBS signals at each pixel point. The sample to be imaged is a solid PDMS with a millimeter-sized pattern of the Tsinghua University logo, fabricated by curing the hydrogel material, as shown in Fig. 3(d). The pattern has a size of 1.26 mm × 0.68 mm and the finest width is approximately 33 µm. The thickness of the PDMS substrate is 3 mm, and that of the protruding pattern is 300 µm. The entire sample is immersed in a methanol solution (99.5%, analytical grade, anhydrous), where both materials are transparent and difficult to distinguish through typical bright-field imaging. While, due to the difference in their viscoelasticity, spatial distribution variations can be investigated without labeling based on mechanical imaging results using the ISBS microscope.

### 3.2 Data processing

The MP method, which is commonly used for analyzing frequency components of exponentially decaying sinusoidal waves in noise, models the time response as a sum of complex exponential decaying sinusoids. Due to its noise suppression capability, the MP method can be applied in ISBS to speed up measurements by reducing the averaging times [23]. When processing ISBS imaging data, since each pixel only shows spectral information of one Brillouin peak, the $M$ parameter in the MP method is set to two. We applied the MP method to the time-domain signals of each pixel using a home-built MATLAB script, allowing three types of spectral information, the Brillouin frequency shift, the linewidth, and the spectral amplitude, to be obtained from a single data processing.

Considering the transition from single-pixel measurement to imaging, the selection of acquisition time can be a concern. For time-domain ISBS signals, an appropriate reduction of the acquisition time contributes to the enhancement of the SNR, as discussed in Sec. II B (3). However, unlike the situation with single-pixel spectral measurements, the acquisition time is constant for each pixel during imaging scans. Therefore, when there is a significant difference in attenuation between various materials in the scanning area, reducing the acquisition time to improve the SNR of the target material (such as the PDMS pattern) may introduce spectral broadening and sidebands to the FFT spectra of some other materials (such as the buffer solution methanol) due to the insufficient acquisition time, thus affecting the accuracy of the linewidth extraction. Unlike traditional FFT-based data processing methods for ISBS signals, the MP method directly estimates the information of each Brillouin peak in the time domain through singular value decomposition, which avoids the spectral distortion introduced by frequency domain transformation, making it immune to issues of spectral broadening and sidebands after FFT for truncated time-domain signals.

We compare the extraction accuracy for the Brillouin peak between FFT and the MP method at short acquisition times through simulation. Fig. 4(a) shows the simulated noise-contaminated time-domain signal of methanol, with a frequency shift set at 265.8 MHz and a linewidth set at 5 MHz, serving as the standard for spectral extraction results. The spectra obtained through

FFT at three different acquisition times are compared as blue dotted lines in Fig. 4(b), from which it can be seen that the spectral broadening and sidebands become more pronounced as the acquisition time decreases. However, when using the MP method to extract Brillouin peaks (orange lines), the values are nearly identical to the set standards (green areas), demonstrating its immunity to the spectral broadening and sidebands caused by the time domain truncation.

Fig. 4(c) further compares the Brillouin linewidth extracted from the methanol signals using FFT-based spectral analysis and the MP method for various acquisition times. For each acquisition time, the signal is generated 10 times to calculate the mean and standard deviation of the extracted linewidths. As the acquisition time decreases, the linewidth extracted from FFT spectra deviates significantly from the standard value, whereas the accuracy of the linewidth extracted by the MP method is hardly affected by the reduction in acquisition time. When the acquisition time decreases to 32 ns, the linewidth extracted from FFT spectra increases to 38 MHz, which is significantly larger than the standard value, while, under the same condition, the linewidth extracted using the MP method remains at 5 MHz. To ensure the accuracy of linewidth extraction while considering its precision, and to maximize the SNR of PDMS by reducing the acquisition time, we consider an acquisition time of about 64 ns to be optimal, where the MP method extracts a linewidth with an accuracy of 0.4 MHz and a precision of 1.7 MHz. To further simulate the experimental conditions, the sampling interval is set to 62.5 ps, and the number of sampling points is 1072. Under this condition, the corresponding spectral SNR for PDMS is 15±1 dB, and the frequency shift deciphered by the MP method has a precision of 1 MHz, while for methanol, the deciphered linewidth is 5.0±1.6 MHz, providing high precision and accuracy.

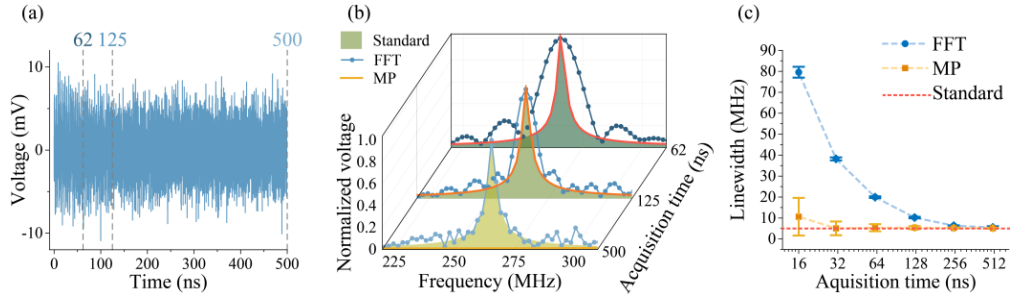

Fig. 4. Immunity of the MP method to spectral broadening and sidebands caused by insufficient acquisition time. (a) Simulated time-domain signal of methanol and (b) corresponding FFT and MP spectra at three acquisition times and the comparison with the standard Brillouin peak. (c) Linewidths at different acquisition times deciphered using the FFT spectra and the MP method, and the comparison with standard values.

## 4. Results and discussion

### 4.1 High-speed Brillouin spectrum measurement

To showcase the efficacy of multi-parameter optimization for SNR, we perform experiments with and without phase compensation and compare the results obtained at various acquisition times for both PDMS and methanol.

On one hand, we acquire time-domain signals of PDMS using setups with and without phase compensation to demonstrate its effectiveness, over 10 repeated measurements, as compared in Fig. 5(a), resulting in the corresponding FFT spectra [Fig. 5(b)]. Since the noise is dominant at an integration time of 0.3 ms without phase compensation, we temporarily extend the integration time to 2.56 ms to substantiate the enhancement in SNR through quantitative comparison. With consistent integration time, laser power, and target sample as described in Sec. III A, the time-domain signal shows a 19-fold increase in amplitude via phase compensation, improving the spectral SNR from 14±1 dB to 21±1 dB. The standard deviation of the frequency shift deciphered by the MP method for PDMS decreases from 3 MHz to 0.2

MHz, and that of the linewidth decreases from 7 MHz to 0.7 MHz, corresponding to a significant improvement in precision achieved through phase compensation. On this basis, the integration time can be further reduced to 0.3 ms and applied to the following experiments.

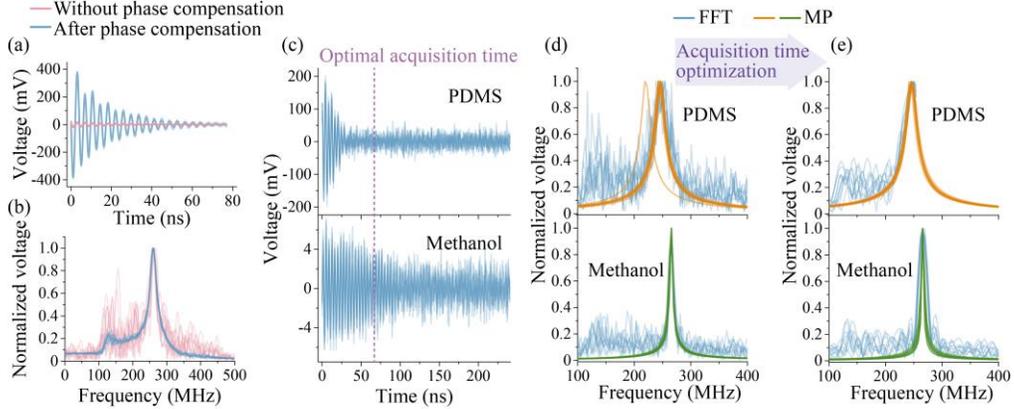

Fig. 5. High-speed ISBS spectroscopy. (a) Comparison of PDMS signals before and after phase compensation and (b) the corresponding FFT spectra. (c) Time-domain signals of PDMS and methanol, and (d) the corresponding spectra at an acquisition time of 241 ns and (e) 67 ns. The blue spectra represent the FFT results, and the orange and green spectra are MP-deciphered Brillouin peaks of PDMS and methanol, respectively.

On the other hand, results under adequate and optimal truncated acquisition times are compared experimentally for both PDMS and methanol. Fig. 5(c) performs the time-domain signals of PDMS and methanol under an integration time of 0.3 ms and an acquisition time of 241 ns, each with 10 repeated measurements, from which it can be seen that the PDMS signal attenuates much faster that of methanol. The purple dashed line represents the optimal acquisition time, 67 ns, based on the simulation results in Secs. II B (3) and III B. The top of Fig. 5(d) and Fig. 5(e) illustrate the PDMS spectra at acquisition times of 241 ns and 67 ns, respectively. For PDMS, a too-long acquisition time reduces the spectral SNR, as noise dominates after optimal acquisition time. The optimization of the acquisition time improves the spectral SNR from 21±7 to 40±10, as well as the relative shift precision by MP from 3% to 0.5%. For the methanol signal with a much slower attenuation, the acquisition time of 67 ns induces spectral broadening and sidebands due to insufficient acquisition, as shown in the bottom of Fig. 5(e). In this case, the linewidths extracted from the FFT spectra are up to 18.9 ± 0.7 MHz, while the MP method helps to keep the deciphered linewidths at 5.0 ± 1.4 MHz without being affected by spectral distortion. Therefore, the acquisition time optimization combined with the MP method maintains high spectral precision in fast ISBS measurements, as well as high spectral resolution considering its minimal impact on linewidth decipher accuracy.

After optimizing the phase compensation, $R_{\text{ref}}$, and acquisition time, the spectral SNRs for PDMS and methanol are measured as 15.6±1.2 dB and 14.5±0.6 dB, respectively, at an integration time of 0.3 ms and the laser power set to the values mentioned in Sec. III A. With further assistance from the MP method, the extracted frequency shift and linewidth for PDMS are 246.0±1.3 MHz and 28.9±1.5 MHz, and for methanol, the corresponding values are 265.8±0.7 MHz and 5.0±1.4 MHz. The results of methanol are utilized to assess the spectral performance, demonstrating a relative shift precision of 0.26% and a relative spectral resolution (linewidth-to-shift ratio) of 1.9%.

*4.2 Three-dimensional mechanical imaging*

High-speed ISBS imaging is demonstrated with the methanol-filled PDMS sample. We perform mechanical imaging in the x-y plane with a single pixel integration time of 0.3 ms. The distribution of Brillouin frequency shift, linewidth, and spectral amplitude for the sample is mapped in Fig. 6(a) with a total pixel number of 140 × 84 and each pixel size of 10 μm, including the corresponding histograms. Even with sub-millisecond integration time, the viscoelastic properties of the two materials can be distinguished. The same data processing method is applied to the acquired signal at each pixel. The orange and green curves in the histograms represent the pixel distributions of PDMS and methanol, respectively, after Gaussian fitting. It is worth noting that when measurements are taken near the boundary between PDMS and methanol, the phase-matching condition is disrupted, resulting in a significant decrease in the spectral SNR. The distribution of these noise-contaminated pixels can also be identified in the histogram of spectral amplitude after phase compensation.

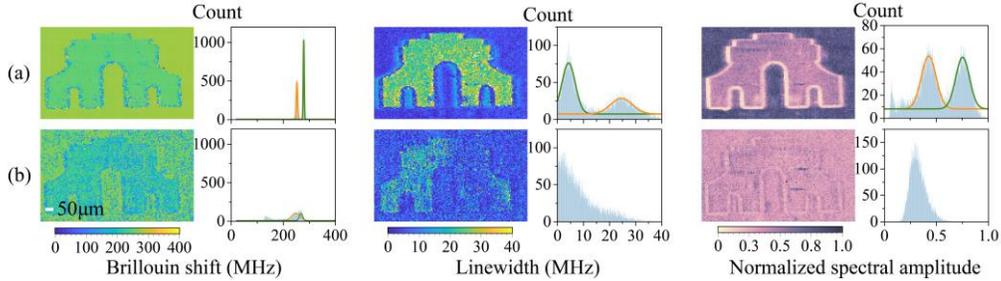

Fig. 6. ISBS imaging at a spectral integration time of 0.3 ms per pixel for the PDMS sample immersed in methanol. (a) The mapping results of Brillouin shift, linewidth, spectral amplitude, and corresponding histograms, with and (b) without phase compensation, where the short white line is a 50 μm scale, and the orange and green curves represent the Gaussian fitting of PDMS and methanol, respectively.

To further highlight the advantages of phase compensation for ISBS imaging, in Fig. 6(b), we present the imaging results and histograms obtained from the ISBS microscope without phase compensation. After applying for phase compensation under the same settings, the image SNR of the Brillouin shift, which is calculated as the ratio of the standard deviation of the maximum pixel count for PDMS and that of the pixel count in the noise region in the histogram, increased from 11.6 dB to 26.2 dB, representing a significant approximately 30-fold improvement. This enhancement renders a more distinct variation on the spatial distributions of Brillouin shift, and thus the elasticity. In addition, the ambiguous distributions of linewidth and spectral amplitude for the two materials, which are associated with viscosity and elasto-optic coefficient [24], respectively, become distinguishable after phase compensation. Despite the small spatial differences of the Brillouin shift due to the similarity in sound speed between PDMS and methanol, the imaging results of linewidth and spectral amplitude exhibit significant spatial variations, primarily attributed to the substantial differences in viscosity and elasto-optic coefficient, enhancing the sensitivity for the characterization of spatial changes of mechanical properties. As a result, the optimization of SNR not only speeds up imaging but also enriches the mechanical information to provide more sensitive mechanical imaging.

On this basis, three-dimensional mapping results with three mechanical contrast mechanisms are shown in Fig. 7(a). The pixel interval in the x-y plane is 10 μm, while the scanning step in the z-axis is 50 μm. The integration time per pixel remains at 0.3 ms. Taking the lower surface of the PDMS pattern as the reference point for the z-axis, as the z-coordinate increases, the mechanical distribution gradually reveals the differences between PDMS and methanol. When the z-coordinate increases to 300 μm, the distribution is fully represented by methanol, which aligns with the known data of the PDMS pattern's thickness. For PDMS and methanol, which exhibit minor differences in Brillouin shifts, the imaging results of linewidth and spectral amplitude contribute to a clearer distinction of mechanical changes, indicating that compared to solely mapping the Brillouin shift, the mapping of all three contrast mechanisms enhances the sensitivity to identify spatial variations.

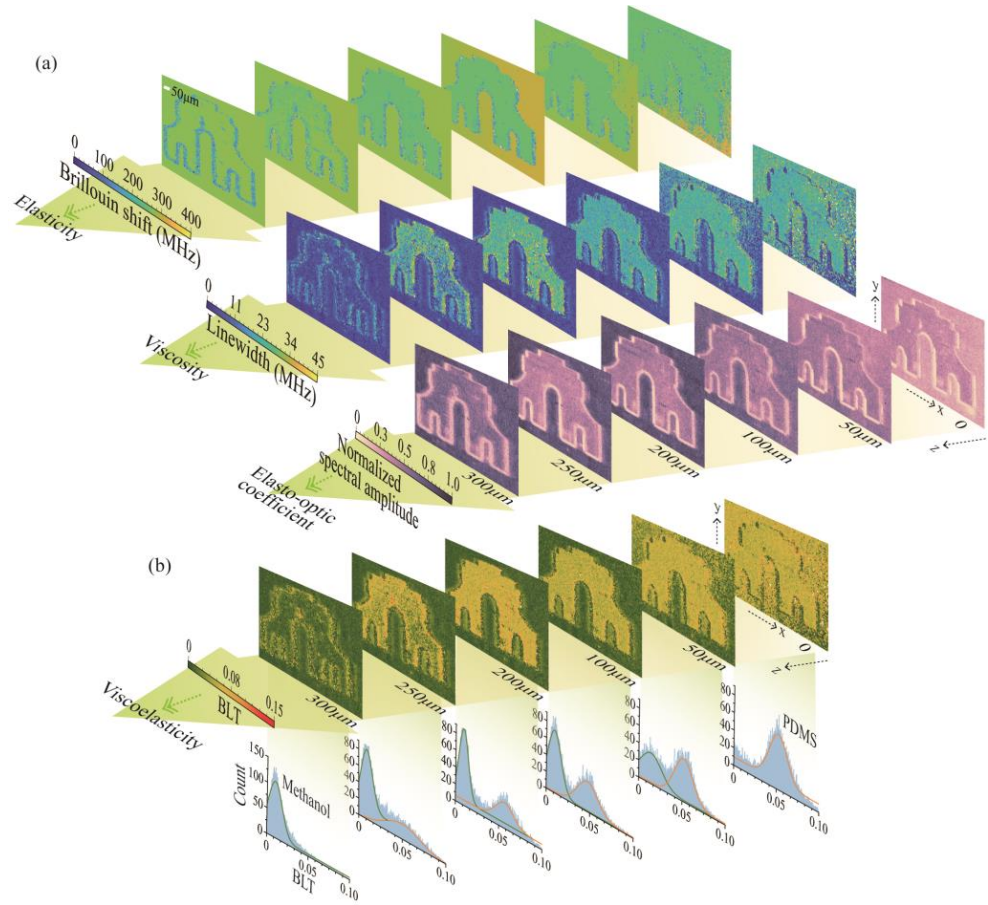

Fig. 7. Three-dimensional imaging results through the high-speed ISBS microscope. (a) Mapping results of Brillouin shift, linewidth, and normalized spectral amplitude. (b) BLT images and corresponding histograms. The orange and green curves represent the BLT distribution of PDMS and methanol, respectively.

To further characterize the longitudinal modulus distribution of the sample, three-dimensional BLT images at six axial positions are mapped in Fig. 7(b), along with the corresponding histograms. According to Eq. (4), BLT is only related to the Brillouin shift and linewidth. Therefore, for materials with unknown density, this method provides a straightforward way to visualize viscoelastic differences without additional measurement. After applying Gaussian fitting to the histograms of BLT, normal distribution curves corresponding to PDMS (orange) and methanol (green) are obtained. The BLT values for methanol at different z-axis positions are mainly distributed around 0.010±0.002, while for PDMS, they are distributed around 0.051±0.003, indicating a higher effective viscosity. For the ISBS imaging system, the introduction of BLT can provide a more intuitive representation of spatial variations in viscoelasticity, offering an additional contrast mechanism to sensitively identify longitudinal modulus changes in time and space.

## 5. Conclusions

For Brillouin microscopy, the limited imaging speed emerges as a pivotal challenge that impedes its wider application in mechanical imaging. Among these, ISBS holds promising capabilities for high-speed imaging but is constrained by factors such as low SNR for single-shot excitation, excessive averaging requirements, and insufficient spatial resolution. After optimizing the spatial resolution of the ISBS microscope to achieve lateral resolution at the level of ten micrometers and axial resolution at the level of a hundred micrometers, we focus on enhancing the SNR of single-shot excitation through phase compensation, reference power optimization, and acquisition time optimization, theoretically grounded in establishing an SNR model for the heterodyne ISBS system. The PDMS sample immersed in methanol is transparent, but its millimeter-scale pattern could still be distinguished by the three-dimensional ISBS imaging results in multiple contrast mechanisms without any contact or labeling. By incorporating the MP method for data processing, the spectral integration time for a single pixel can be reduced to 0.3 ms under an energy density of 0.001 nJ/μm$^3$ at 532 nm and 1 nJ/μm$^3$ at 780 nm, without obvious photodamage, while maintaining a relative spectral precision of 0.26% and a relative spectral resolution of 1.9%. Building upon this, the BLT parameter is introduced as an additional ISBS contrast mechanism to concisely characterize spatial and temporal variations of longitudinal modulus.

A sub-millisecond single-pixel integration time offers the possibility to capture rapid biomechanical changes. The current spatial resolution essentially meets the requirements of biomedical imaging, but its further improvement is constrained by the acoustic wavelength scale. For both spontaneous and CW-stimulated schemes, the acoustic wave propagation is parallel to the optical axis, and the measurement of Brillouin spectra is typically based on the backscattered lights. In this situation, the interacting acoustic wavelengths are comparatively short. While in the ISBS scheme, acoustic wave propagation is perpendicular to the optical axis, and forward-scattered light is acquired, leading to a more significant limitation on spatial resolution due to longer acoustic wavelengths, which can be alleviated by increasing the angle between the two pump beams. However, the increase in acoustic frequency due to the decrease in acoustic wavelength would significantly raise the bandwidth requirements for the detection system. In subsequent research, after reducing the acoustic wavelength via a TG with a shorter period and a 4*f* system with a higher amplification rate, an acoustic-optical modulator can be implemented in the reference beam to decrease the frequency of the ISBS signal, thus mitigating the bandwidth constraints on the detection system. Through the enhancement of the spatial resolution to the micron level and the integration of a spectral measurement speed of 10 kHz, the ISBS microscope has the potential to promote the investigation of mechanical changes in tissues and cells at finer time scales.

**Funding.** National Key Research and Development Program of China (2020YFC2200101).

**Disclosures.** The authors declare no conflicts of interest.

**Data availability.** Data underlying the results presented in this paper are not publicly available at this time but may be obtained from the authors upon reasonable request.